\documentclass[reprint,amsmath,amssymb,superscriptaddress]{revtex4-2}

\usepackage{bm}
\usepackage{graphicx}
\usepackage{amsmath}
\usepackage{amsbsy}
\usepackage{amstext}
\usepackage{amssymb}
\usepackage{units}
\usepackage[utf8]{inputenc}
\usepackage[english]{babel}
\usepackage[colorlinks, citecolor=blue, urlcolor=blue, linkcolor=red,breaklinks]{hyperref}
\usepackage{gensymb}

\begin{document}
	
\title{Coulomb drag in metallic twisted bilayer graphene}
	
\author{Federico Escudero}
\email{federico.escudero@uns.edu.ar}
\affiliation{Departamento de Física, Universidad Nacional del Sur, Av. Alem 1253, B8000, Bahía Blanca, Argentina}
\affiliation{Instituto de Física del Sur, Conicet, Av. Alem 1253, B8000, Bahía Blanca, Argentina}
\author{Juan Sebastián Ardenghi}
\affiliation{Departamento de Física, Universidad Nacional del Sur, Av. Alem 1253, B8000, Bahía Blanca, Argentina}
\affiliation{Instituto de Física del Sur, Conicet, Av. Alem 1253, B8000, Bahía Blanca, Argentina}
	
\begin{abstract}
Strongly correlated phases in twisted bilayer graphene (TBG) typically arise as transitions from a state in which the system behaves as a normal metal. In such metallic regime, electron-electron interactions usually only play a subleading role in transport measurements, compared to the dominant scattering mechanism. 
Here, we propose and theoretically study an exception to this based on a Coulomb drag setup between two metallic TBG, separated so that they only couple through many-body interactions. We find that by solely varying the twist angle equally in both TBG, the drag resistivity exhibits a unique maximum as the system crossovers from a degenerate to a nondegenerate regime. When the twist angles in each TBG differ, we find an anomalous drag resistivity characterized by the appearance of multiple peaks. We show that this behavior can be related to the dependence of the rectification function on the twist angle.
\end{abstract}

\maketitle  

\section{Introduction}

Over the last years, several experiments have reported the existence of numerous correlated phases in twisted bilayer graphene (TBG) around a \textit{magic} angle $\theta_M\sim 1.05{^\circ}$, such as unconventional superconductivity and metal-insulator transitions \cite{Cao2018,Cao2018a,Yankowitz2019,Wong2020,Oh2021,Pierce2021}. These phenomena are thought to arise from electronic interactions that are greatly enhanced as the bands become flat at the magic angle \cite{Santos2007,Shallcross2010,Luican2011,Bistritzer2011,Tarnopolsky2019}. The relation of these interaction with external parameters, such as the temperature or the carrier density, determines the state of the system. Consequently, much effort has been made to elucidate the transport properties of TBG due to many-body interactions \cite{Nimbalkar2020,Andrei2020,Liu2021}. A particular but relevant case is the normal metal state of TBG, which usually occurs at temperatures above which the superconductivity is observed \cite{Wagner2022,Chen2020}. The study of many-body effects in metallic TBG may thus help to understand the origin and nature of the correlated phases. However, although it is clear that electron-electron interactions play a major role in the rich phase diagram of TBG, their role within its normal metallic state is less evident. In part this is because in such regime many-body interactions often only play a subleading role in transport measurements, compared to the dominant scattering mechanism \cite{Chung2018,Andelkovic2018,Polshyn2019,Wu2019}. 

In this work we propose a direct method to study many-body interactions within the metallic regime, based on a Coulomb drag effect between two TBG that are closely spaced, but such that no interlayer hopping between them is possible. Both TBG thus only couple through long-range Coulomb interactions. The drag effect arises when an external electric current driven in one layer induces a voltage difference in other closely spaced layer \cite{Rojo1999,Narozhny2016}. Typically, such effect depends directly on the many-body interlayer interactions; the main mechanism is the Coulomb interaction, but phonon-mediated or photon-mediated interactions may also contribute, especially at large interlayer separations \cite{Gramila1993,Boensager1998,Berman2010,Escudero2022}. The drag resistivity, i.e., the ratio between the applied current in the active layer and the voltage induced in the passive layer, reflects the response of the system due to the interlayer interactions, as well as the temperature and carrier density in each layer \cite{Narozhny2012}. Thus, Coulomb drag measurements between two metallic TBG may allow one to elucidate properties of the electron-electron interactions in the system, to a degree that is not directly available in transport measurements carried out over a single TBG.

Here, we particularly focus on the drag at low temperatures and carrier densities, where the transport in metallic TBG is dominated by disorder and phonons \cite{Yudhistira2019,Sharma2021}. We find that the drag resistivity $\rho_D$ depends strongly on the angle-dependent Fermi velocity $v^{\star}$ in three general aspects: (i) the renormalization of the coupling constant $\alpha^{\star}\propto 1/v^{\star}$; (ii) the relation between the chemical potential $\mu$ and the temperature $T$, which in turn determines the regime of the system; and (iii) the interplay between intraband ($\omega<v^{\star}q$) and interband ($\omega>v^{\star}q$) scattering. When both TBG have the same twist angle, the drag effect follows a conventional behavior in which, as the system crossovers from a degenerate to a nondegenerate regime, the drag resistivity peaks
around $\mu/k_BT\sim 2$. However, when the twist angles are different we find that the drag resistivity follows a nontrivial behavior, characterized by the appearance of several peaks. The shape of these peaks depends strongly on the twist difference, as well as the temperature, carrier density, and distance between the TBG. A qualitative explanation is given in terms of the angle-dependence of the nonlinear susceptibility.

This work is organized as follows: In Sec. \ref{sec:TheoreticalModel} we describe the theoretical model used to study the Coulomb drag between two metallic TBG, both of which are described by a two-band model within the Dirac approximation. Semi-analytical expressions are obtained in order to compute the nonlinear susceptibility, taking into account the full energy-dependence of the scattering time in TBG, with contributions of both gauge phonons and charged impurities. The dynamically screened Coulomb interaction is obtained within the random phase approximation. In Sec. \ref{sec:ResultsDiscussion} we present and discuss the numerical results for the drag resistivity, in the cases of equal twist angle in both TBG, and different twist angles. In the latter case we provide an intuitive explanation for the observed anomalous drag behavior, based on how the product of two nonlinear susceptibilities changes depending on the difference between the twist angles. Finally, our conclusion follow in Sec. \ref{sec:Conclusions}. 

\section{Theoretical model}\label{sec:TheoreticalModel}

\subsection{Proposed setup}

\begin{figure}[t]
	\includegraphics[scale=0.6]{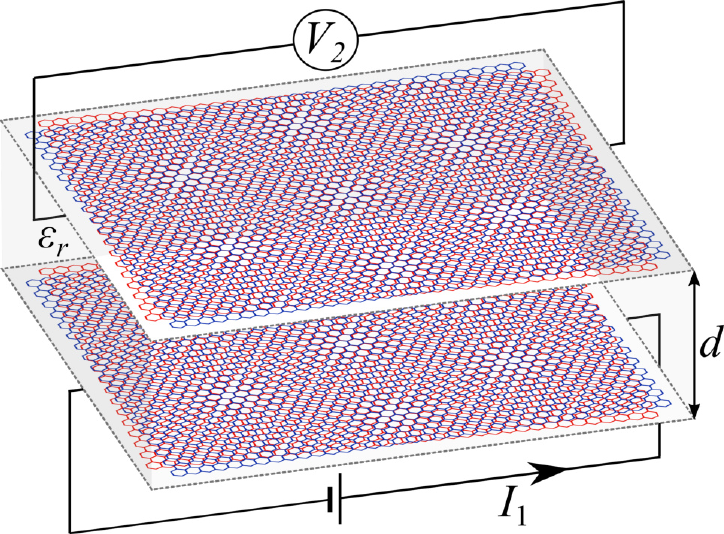}	
	\caption{Schematic setup to study the Coulomb drag between two twisted bilayer graphene (TBG). The TBG are assumed to be separated by a dielectric medium by a distance $d$, so that no tunneling between is possible. Electrons on each TBG can only interact through long-range many-body interactions. In a drag configuration, these interactions can induce a voltage difference $V_2$ in one TBG if a current $I_1$ driven in the other TBG. The drag resistivity is determined by the ratio $\rho_D\propto V_2/I_1$. }\label{fig:SchmaticDrag}
\end{figure}

The schematic drag setup is shown in Fig. \ref{fig:SchmaticDrag}. The two TBG are separated so that no tunneling between them is possible, and they interact with one another only through long-range Coulomb interactions. These interactions can induce a voltage in one TBG (referred as \textit{passive} TBG) if a current is driven through the other TBG (referred as \textit{active} TBG). Throughout this work we assume that the twist angle in each bilayer graphene can be varied independently. Although this experimental configuration has not yet been realized, it seems feasible given the recent advances in fabricating moiré heterostructures \cite{Cai2021,Yang2022,Zhang2021,Liu2021a,Cao2021,Kennes2021}.      

The leading order contribution to the drag conductivity $\sigma_D$ can be calculated using either the diagrammatic approach \cite{Kamenev1995,Flensberg1995a,Narozhny2012}, or the kinetic theory approach \cite{Jauho1993,Hwang2011,Lux2012,Escudero2022a}. For a homogeneous system at a uniform temperature $T$ one gets
\begin{equation}
	\sigma_{D}=\frac{e^{2}}{16\pi k_{B}T}\sum_{\mathbf{q}}\int_{-\infty}^{\infty}d\omega\left|U_{12}\right|^{2}\frac{\Gamma_{1}\Gamma_{2}}{\sinh^{2}\left(\hbar\omega/2k_{B}T\right)}.\label{eq:dragcond}
\end{equation}
Here $U_{12}\left(\mathbf{q},\omega\right)$ is the dynamically screened interlayer Coulomb
interaction, and $\Gamma_{\ell}$ is the nonlinear susceptibility (NLS) in the $\ell=1,2$ TBG, projected along the current direction in the active TBG. From $\sigma_D$, the drag resistivity is obtained by inverting the $2\times2$ conductivity matrix, $\rho_{D}\simeq-\sigma_{D}/\sigma_1\sigma_2$, where $\sigma_{\ell}$ ($\gg\sigma_D$) is the conductivity within each TBG. 

\subsection{Two-band Dirac model of TBG}\label{subsec:TwoBand}

In this work we restrict our analysis to the drag between two TBG
in the metallic regime. For low carrier densities, the electronic
properties of metallic TBG are well captured by a two-band model in
which electrons behave as massless chiral fermions with a Dirac-like Hamiltonian \cite{Santos2007,Shallcross2010,Bistritzer2011}
\begin{equation}
	\hat{H}_{0,\ell}=-i\hbar v^{\star}_{\ell}\int d\mathbf{r}\hat{\psi}_{\ell}^{\dagger}\left(\mathbf{r}\right)\boldsymbol{\sigma}\cdot\nabla\hat{\psi}_{\ell}\left(\mathbf{r}\right),\label{eq:H0}
\end{equation}
with a renormalized Fermi-velocity \cite{Bernevig2021a}
\begin{equation}
	v_{\ell}^{\star}=v\frac{1-3\alpha_{\ell}^{2}}{1+6\gamma^{2}\alpha_{\ell}^{2}},\label{eq:vrenorm}
\end{equation}
where $v$ is the Fermi velocity in monolayer
graphene \cite{CastroNeto2009}. Here $\alpha_{\ell}=w_{1}/\hbar vk_{\theta,\ell}$ and $2\gamma^{2}=1+\left(w_{0}/w_{1}\right)^{2}$,
where $w_{1}\simeq0.11$ eV and $w_{0}\simeq0.8w_{1}$ are the hopping
energies of AB/BA and AA stacking, respectively \cite{Jain2016,Koshino2020}, while $k_{\theta,\ell}=8\pi\sin\left(\theta_{\ell}/2\right)/3a$
is the wave vector magnitude of the moiré Brillouin zone ($a\simeq2.46\,\textrm{Å}$
is the lattice constant in graphene). For low twist angles, close
to the first magic angle $\theta_{M}\sim1.05^{\circ}$ (but not exactly
at), the two-band model remains a good approximation for carrier densities
$n\lesssim10^{11}\;\mathrm{cm^{-2}}$ \cite{Yudhistira2019}.

The field operators can be expanded in momentum space as
$\hat{\psi}_{\ell}\left(\mathbf{r}\right)=A^{-1/2}\sum_{\mathbf{k},s}e^{i\mathbf{k}\cdot\mathbf{r}}\hat{c}_{\ell,\mathbf{k},s}u_{\mathbf{k},s}$,
where $A$ is the area of the system. The operator $\hat{c}_{\ell,\mathbf{k},s}$
creates an electron in the $\ell=1,2$ TBG, with momentum $\mathbf{k}$
in the $s=\pm$ band. The Dirac approximation requires $\left|\mathbf{k}\right|\ll k_{\theta}$.
The pseudospinor $u_{\mathbf{k},s}^{\dagger}=\left(\begin{array}{cc}
	1 & se^{-i\varphi_{\mathbf{k}}}\end{array}\right)/\sqrt{2}$, where $\tan\varphi_{\mathbf{k}}=k_{y}/k_{x}$, comes from the sublattice
structure in graphene \cite{CastroNeto2009,DasSarma2011}.
Replacing in Eq. \eqref{eq:H0} leads to the energy operator $\hat{H}_{0,\ell}=\sum_{\mathbf{k},s}\epsilon_{\ell,\mathbf{k},s}\hat{c}_{\ell,\mathbf{k},s}^{\dagger}\hat{c}_{\ell,\mathbf{k},s}$,
where $\epsilon_{\ell,\mathbf{k},s}=s\hbar v_{\ell}^{\star}\left|\mathbf{k}\right|$
is the dispersion relation in the $s$ band. The pseudospinor $u_{\mathbf{k},s}$ yields the well-known chirality factor within the Dirac approximation,
\begin{equation}
	F_{ss'}\left(\mathbf{k},\mathbf{q}\right)=\left|u_{\mathbf{k}+\mathbf{q},s'}^{\dagger}u_{\mathbf{k},s}\right|^{2}=\frac{1+ss'\cos\left(\varphi_{\mathbf{k}}-\varphi_{\mathbf{k+q}}\right)}{2}.
\end{equation}

\subsection{Scattering time and conductivity}

The conductivity $\sigma_{\ell}$ depends on the dominant scattering mechanism. In metallic TBG, at low temperatures this scattering has, in general, non-negligible contribution from impurities and phonons \cite{Wu2019,Zarenia2020,Sharma2021}. The latter comes from gauge phonons that are immune to the strong screening that arises in the flat bands of TBG around the magic angle \cite{Yudhistira2019}. 

For the impurity scattering we consider long-range Coulomb disorder, taking into account the static screening within the random phase approximation \cite{Hwang2009}
\begin{equation}
\frac{1}{\tau_{\mathrm{i,{\ell}}}}=2\pi\left(\alpha^{\star}_{\ell}\right)^{2}n_{i}v^{\star}_{\ell}k\int_{0}^{\pi}d\theta\frac{1-\cos^{2}\theta}{\left[2k\sin\left(\theta/2\right)+q_{T,\ell}\right]^{2}}.\label{eq:ScattImp}
\end{equation}
Here $\alpha^{\star}_{\ell}=e^{2}/4\pi\hbar\varepsilon_{0}\varepsilon_{r}v^{\star}_{\ell}$
is the coupling constant, $n_{i}$ is the impurity density, and $q_{T,{\ell}}=q_{0T,{\ell}}\tilde{\Pi}_{\ell}\left(q,T\right)$
is the screened Thomas-Fermi momentum, where $q_{0T,{\ell}}=g\alpha^{\star}_{\ell}k_{F,\ell}$
($g=8$ is the flavor degeneracy in TBG), and $\tilde{\Pi}_{\ell}\left[q=2k\sin\left(\theta/2\right),T\right]$
is the static polarization at finite temperatures \cite{DasSarma2011}. For the phonon scattering, since
we will restrict our analysis to twist angles for which the Fermi velocity is still much higher than the phonon velocities, we take
into account only the intraband scattering \cite{Wu2019,Sharma2021}. For gauge phonons one then has \cite{Yudhistira2019,Sharma2021}
\begin{equation}
\frac{1}{\tau_{\mathrm{p},\ell}}=\sum_{\nu}\frac{\zeta_\ell^{2}}{2h\rho c_{\nu}v^{\star}_{\ell}}\int_{0}^{2k}dq\frac{q^{3}}{k^2}\sqrt{1-\left(\frac{q}{2k}\right)^{2}}\left(1+2n_{\mathbf{\ell,q},\nu}\right),\label{eq:ScattPhon}
\end{equation}
where the summation is over the two acoustic phonon branches (LA and TA), for which we assume an average velocity
$c_{\nu}\simeq2\times10^{6}\,\mathrm{cm/s}$ (independent of the twist
angle). Here $\zeta_\ell=\beta_{A}v^{\star}_\ell/v\tan\left(\theta/2\right)$
is the coupling constant \cite{Lian2019} ($\beta_{A}\sim3.6$ eV), $\rho$ is
the mass density in monolayer graphene, and $n_{\ell,\mathbf{q},\nu}$
is the Bose-Einstein distribution. 

\begin{figure}[t]
	\includegraphics[scale=0.35]{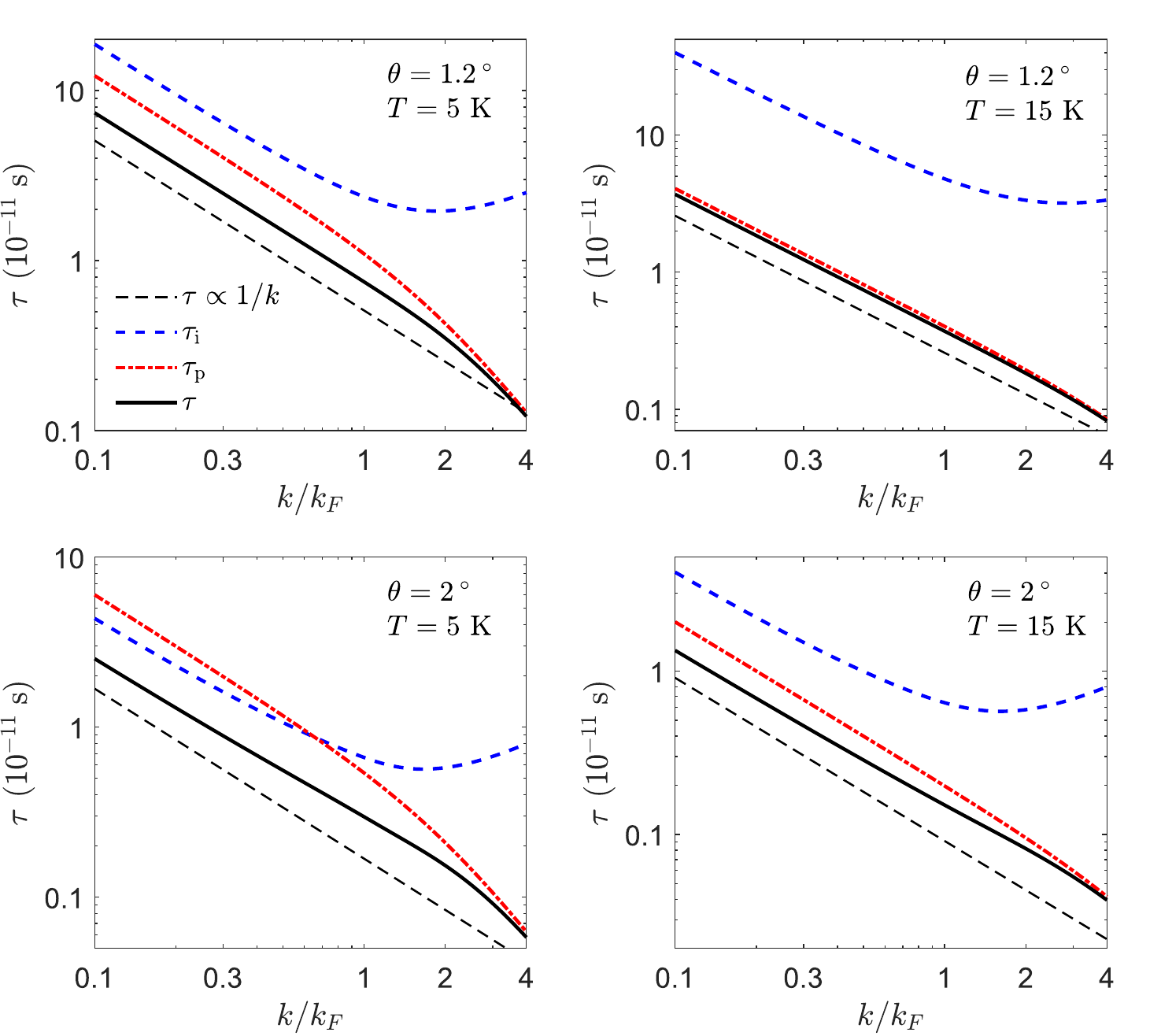}	
	\caption{Momentum dependence of the scattering time (ST) in metallic TBG, for different twist angles and temperatures. In blue dashed-line is shown the ST $\tau_{\mathrm{i}}$ due to long-range screened charged impurities with density $n_i=10^{10}\;\mathrm{cm^{-2}}$, in red dot-dashed line the ST $\tau_{\mathrm{p}}$ due to gauge phonons, and in solid black line the net ST $\tau^{-1}=\tau_{\mathrm{i}}^{-1}+\tau_{\mathrm{p}}^{-1}$. In black dashed-line is schematically shown a dependence $\tau\propto1/k$.}\label{fig:Scattering}
\end{figure}

Using the Matthiessen's rule, which is a good approximation at relatively large twist angles \cite{Hwang2020}, the net scattering time is given
by $\tau_{\ell}^{-1}=\tau_{\mathrm{i,\ell}}^{-1}+\tau_{\mathrm{p,\ell}}^{-1}$. Unless the temperature is very low, and the twist angle is relatively large, $\tau^{-1}_\ell$ is mostly determined by the phonon contribution, roughly yielding a momentum dependence $\tau_\ell\propto1/k$ \cite{Sharma2021,Yudhistira2019} (see Fig. \ref{fig:Scattering}). 

Given the scattering time, the conductivity is then calculated as $\sigma_\ell=ge^{2}\left(v_{\ell}^{\star}\right)^2\sum_{\mathbf{k},s}\tau_{\ell}\left(-\partial f_{\ell,\mathbf{k},s}/\partial\epsilon_{\ell,\mathbf{k},s}\right)$,
where $f_{\ell,\mathbf{k},s}$ is the Fermi-Dirac distribution. The chemical potential $\mu_\ell$ is obtained numerically from the equation for the carrier density, $n_\ell=\left(g/A\right)\sum_{\mathbf{k},s}\left[f_{\ell,\mathbf{k},s}\left(\mu_\ell\right)-f_{\ell,\mathbf{k},s}\left(\mu_\ell=0\right)\right]$, which implies
\begin{equation}
	n_\ell=\frac{g}{2\pi}\left(\frac{k_{B}T}{\hbar v^{\star}_\ell}\right)^{2}\left[\mathrm{Li}_{2}\left(-e^{-\beta\mu_\ell}\right)-\mathrm{Li}_{2}\left(-e^{\beta\mu_\ell}\right)\right],\label{eq:carrierdensity}
\end{equation}
where $\mathrm{Li}_{2}\left(x\right)$ is the dilogarithm function.

\subsection{Nonlinear susceptibility}

Within the two-band model of TBG, the nonlinear susceptibility is
calculated as \cite{Amorim2012,Carrega2012,Escudero2022a}
\begin{align}
	\boldsymbol{\Gamma}_{\ell}\left(\mathbf{q},\omega\right) & =-2\pi g\sum_{\mathbf{k},s,s'}\left(f_{\ell,\mathbf{k},s}-f_{\ell,\mathbf{k}+\mathbf{q},s'}\right)F_{ss'}\left(\mathbf{k},\mathbf{q}\right)\nonumber \\
	& \qquad\qquad\times\left(\tau_{\ell,\mathbf{k}}\mathbf{v}_{\ell,\mathbf{k},s}-\tau_{\ell,\mathbf{k}+\mathbf{q}}\mathbf{v}_{\ell,\mathbf{k}+\mathbf{q},s'}\right)\nonumber \\
	& \qquad\qquad\times\delta\left(\hbar\omega+\epsilon_{\ell,\mathbf{k},s}-\epsilon_{\ell,\mathbf{k}+\mathbf{q},s'}\right),\label{eq:NLS0}
\end{align}
where $\mathbf{v}_{\ell,\mathbf{k},s}=\hbar^{-1}\nabla\epsilon_{\ell,\mathbf{k},s}$ is the velocity vector. 
The NLS at finite temperatures, beyond the degenerate regime, is typically obtained by assuming a constant scattering time \cite{Badalyan2012,Narozhny2012,Fandan2019}. Since this would not capture the strong momentum-dependence of $\tau_{\ell,\mathbf{k}}$ in metallic TBG, we compute the NLS semi-analytically by rather considering a scattering time of the form $\tau_{\ell}\left(\left|\mathbf{k}\right|=k\right)$,
only imposing the restriction that it is isotropic \cite{Carrega2012}.
After straightforward algebraic manipulations we then find the general expression (see appendix \ref{app:NLS})
\begin{align}
	\boldsymbol{\Gamma}_{\ell}\left(\mathbf{q},\omega>0\right) & =\frac{g}{4\pi\hbar}\frac{1}{\sqrt{\left|q^{2}-\omega_{\ell}^{2}\right|}}\frac{\mathbf{q}}{q^{2}}\left[\Theta\left(q-\omega_{\ell}\right)\Gamma_{\ell,+}\left(q,\omega\right)\right.\nonumber \\
	& \qquad+\left.\Theta\left(\omega_{\ell}-q\right)\Gamma_{\ell,-}\left(q,\omega\right)\right],\label{eq:NLS}
\end{align}
where $\Theta$ is the step function, $\omega_{\ell}=\omega/v^{\star}_{\ell}$ and
\begin{align}
	\Gamma_{\ell,+}\left(q,\omega\right) & =\sum_{s=\pm1}\int_{\left(q-s\omega_{\ell}\right)/2}^{\infty}dk\mathcal{F}_{\ell,s}\left(k,\omega\right)\mathcal{K}_{\ell,s}\left(k,q,\omega\right),\nonumber \\
	\Gamma_{\ell,-}\left(q,\omega\right) & =\int_{\left(\omega_{\ell}-q\right)/2}^{\left(\omega_{\ell}+q\right)/2}dk\mathcal{F}_{\ell,-1}\left(k,\omega\right)\mathcal{K}_{\ell,-1}\left(k,q,\omega\right),
\end{align}
with
\begin{align}
	\mathcal{F}_{\ell,s}\left(k,\omega\right) & =f_{\ell}\left(k\right)+f_{\ell}\left(-k\right)-f_{\ell}\left(k+s\omega_{\ell}\right)\nonumber \\
	& \qquad-f_{\ell}\left(-k-s\omega_{\ell}\right),\\
	\mathcal{K}_{\ell,s}\left(k,q,\omega\right) & =\frac{s}{k}\sqrt{\left|\left(\omega_{\ell}+2sk\right)^{2}-q^{2}\right|}\nonumber \\
	& \qquad\times \tau_{\ell}\left(k\right)\left(q^{2}-\omega_{\ell}^{2}-2sk\omega_{\ell}\right).
\end{align}
Semi-analytical expressions for the NLS with an arbitrary energy-dependent scattering time were first obtained in Ref. \cite{Narozhny2012}. In the present case of TBG, the NLS are calculated numerically at finite temperatures by using the full energy and temperature dependence of the scattering time, as given by Eqs. \eqref{eq:ScattImp} and \eqref{eq:ScattPhon}. We note that, in final expression of the drag resistivity,
the divergences in Eq. \eqref{eq:NLS} when $q\rightarrow\omega_{\ell}$  (which
comes from the fact that the dispersion relation is linear \cite{Narozhny2016}) are cured when one takes into consideration the full
dynamical screening of the interlayer interaction \cite{Gangadharaiah2008,Carrega2012}  (see Appendix \ref{app:Screening}).

\subsection{Many-body interactions}

The electron-electron interactions in the proposed setup are described by the Hamiltonian
\begin{equation}
	\hat{H}_{\ell\ell'}=\frac{1}{2}\int d\mathbf{x}d\mathbf{y}\hat{\psi}_{\ell}^{\dagger}\left(\mathbf{x}\right)\hat{\psi}_{\ell'}^{\dagger}\left(\mathbf{y}\right)V_{\ell\ell'}\left(\mathbf{x}-\mathbf{y}\right)\hat{\psi}_{\ell'}\left(\mathbf{y}\right)\hat{\psi}_{\ell}\left(\mathbf{x}\right),\label{eq:Int}
\end{equation}
where 
$V_{\ell\ell'}\left(\mathbf{r}\right)=\left(e^{2}/4\pi\varepsilon_{0}\varepsilon_{r}\right)/\left[\mathbf{r}^{2}+\left(1-\delta_{\ell\ell'}\right)d^{2}\right]$ is the bare Coulomb potential. In principle, the interaction \eqref{eq:Int} can be treated perturbatively if the coupling constant is small \cite{Kamenev1995,Narozhny2012}. For TGB embedded in a homogeneous dielectric medium of relative permitting $\varepsilon_{r}$,
the coupling constant reads 
$\alpha^{\star}_{\ell}\sim\alpha_{g}\left(v/v^{\star}_{\ell}\right)$, where $\alpha_g\sim2.2/\varepsilon_{r}$ is the coupling constant in monolayer graphene \cite{CastroNeto2009}. The value of $\alpha_g$ is expected to depend also on the quasiparticle renormalization due to intralayer many-body interactions, which tend to lower $\alpha_g$ \cite{Kotov2012}. In what follows we set a bare value $\alpha_g=0.3$, which roughly corresponds to $\alpha_g\sim1.3/\varepsilon_{r}$ and a uniform boron nitride dielectric medium ($\varepsilon_{r}\sim4$) \cite{Peres2011}. Since the resulting coupling constant in TBG then becomes of the order of unity already at relatively low velocity renormalizations, we restrict our analysis to $\theta\gtrsim1.2^{\circ}$. Although $\alpha^{\star}_{\ell}$ at low twist angles is clearly not small in the context of QED, it is still within the range found in most metals \cite{Mahan2000}.

The dynamical screening is then calculated within the random phase
approximation (RPA) by coupling the passive and active TBG with a
diagonal polarization matrix \cite{Kamenev1995}. Assuming a drag setup with a homogeneous
dielectric medium \cite{Katsnelson2011,Badalyan2012}, the RPA yields the screened interlayer interaction $U_{12}\left(\mathbf{q},\omega\right)=V_{12}\left(q\right)/\varepsilon_{12}\left(\mathbf{q},\omega\right)$ with
\begin{align}
	\varepsilon_{12}\left(\mathbf{q},\omega\right) & =\left[1+V_{11}\left(q\right)\Pi_{1}\left(\mathbf{q},\omega\right)\right]\left[1+V_{22}\left(q\right)\Pi_{2}\left(\mathbf{q},\omega\right)\right]\nonumber \\
	& \qquad-V_{12}^2\left(q\right)\Pi_{1}\left(\mathbf{q},\omega\right)\Pi_{2}\left(\mathbf{q},\omega\right),\label{eq:dielectricF}
\end{align}
where $V_{\ell\ell'}\left(q\right)=\left(e^{2}/2\varepsilon_{0}\varepsilon_{r}\right)q^{-1}\exp\left[-qd\left(1-\delta_{\ell\ell'}\right)\right]$
is the Fourier transform of the Coulomb potential, and
$\Pi_{\ell}\left(q,\omega\right)$ is the dynamical polarization
in the $\ell$ TBG, 
\begin{equation}
	\Pi_{\ell}\left(\mathbf{q},\omega\right)=-g\sum_{\mathbf{k},s,s'}\frac{\left(f_{\ell,\mathbf{k},s}-f_{\ell,\mathbf{k}+\mathbf{q},s'}\right)F_{ss'}\left(\mathbf{k},\mathbf{q}\right)}{\hbar\omega+\epsilon_{\ell,\mathbf{k},s}-\epsilon_{\ell,\mathbf{k}+\mathbf{q},s'}+i0^{+}}.\label{eq:pol}
\end{equation}
The polarization function gives the dependence of the screened interaction $U_{12}$ on the twist angle in each TBG. Since electrons in TBG can crossover
from a degenerate to a nongenerate regime as the twist angle decreases, it is essential to consider the dynamical screening at finite temperatures \cite{Narozhny2012}. This naturally captures the role of
plasmons in the drag, which are expected to become relevant when $T\gtrsim0.2T_{F}$
\cite{Flensberg1995,Narozhny2012,Badalyan2012}, where $T_F$ is the Fermi temperature. We
compute $\Pi_{\ell}\left(\mathbf{q},\omega\right)$ numerically, at finite temperatures, by using the semi-analytical
expressions of Ref. \cite{Ramezanali2009} for the
polarization operator in monolayer graphene, and taking into account the twist angle by its leading order renormalization of the Fermi
velocity (Appendix \ref{app:Screening}). 

\section{Results and discussion}\label{sec:ResultsDiscussion} 

\subsection{Equal twist angles}

\begin{figure}[t]
	\includegraphics[scale=0.35]{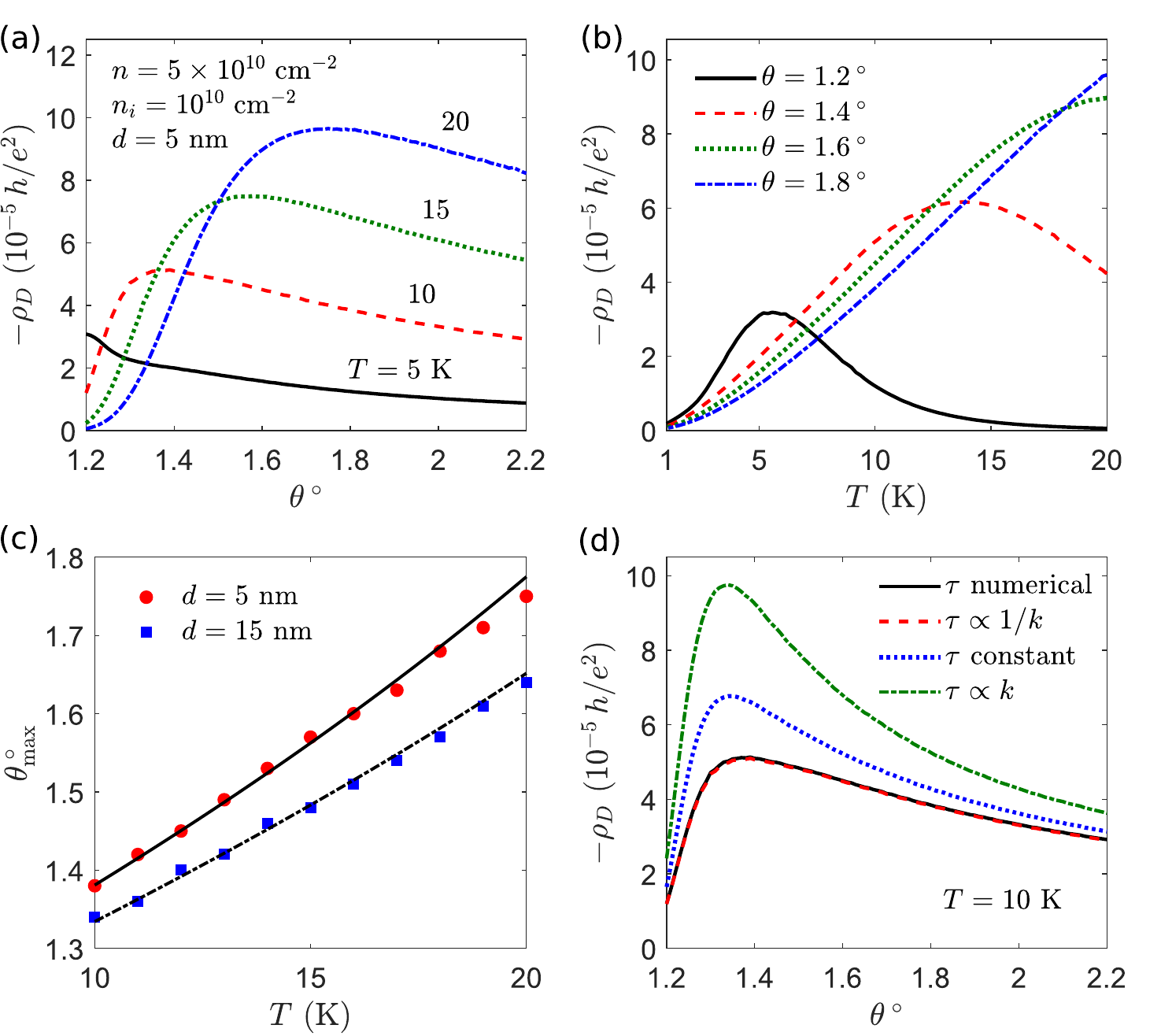}	
	\caption{(a)--(b) Drag resistivity for equal twist angles. The drag peaks as TBG crossovers from a degenerate to a nondegenerate regime. This can take place by solely lowering the twist angle. (c) Twist angle $\theta_{\mathrm{max}}$ at which the drag resistivity is maximum; dots are numerical results, and the lines are obtained from Eq. \eqref{eq:tmax} by fixing the ratio $\beta\mu$ at 3.4 ($d=5$ nm) and 2.8 ($d=15$ nm), around which $\rho_D$ peaks in each case. (d) Drag resistivity for different scattering times $\tau$. The black solid line is the full numerical calculation using Eqs. \eqref{eq:ScattImp} and \eqref{eq:ScattPhon}.}\label{fig:DragEqual}
\end{figure}

Fig. \ref{fig:DragEqual} shows the drag resistivity between two TBG with the same twist
angle $\theta$. The observed behavior can be well explained by a crossover of the system from a degenerate to a nondegenerate regime \cite{Narozhny2012,Lux2012,Badalyan2012,Chen2015}. Indeed, since the chemical potential $\mu$ increases as the Fermi velocity in each TBG increases, the ratio $\mu/k_{B}T$ scales with the twist angle, thus leading to the observed drag behavior. As small variations in $\theta$ can lead to relatively large changes in the Fermi velocity, the drag effect
is highly sensitive to the exact twist angle in TBG. For instance,
at $T=15$ K a twist decrease $\Delta\theta\sim0.3^{\circ}$ already
reduces $\rho_{D}$ from its peak by more than one order of magnitude.
Such reduction becomes even more pronounced at higher temperatures. A similar high sensibility to the twist angle is already seen in the resistivity at each TBG \cite{Yudhistira2019,Hwang2020}. 

The peak of the drag resistivity generally occurs when $\beta\mu$
is of the order of unity \cite{Narozhny2012,Lux2012}.
The ratios $\beta\mu$ and $v^{\star}/v$ can be related through
the carrier density equation \eqref{eq:carrierdensity}. Replacing the renormalized velocity \eqref{eq:vrenorm} and solving for the twist angle yields
\begin{equation}
	\theta\simeq\frac{\sqrt{3}w_{1}}{\hbar vk_{D}}\sqrt{\frac{1+2\gamma^{2}F\left(\beta\mu\right)T/T_{Fg}}{1-F\left(\beta\mu\right)T/T_{Fg}}},\label{eq:tmax}
\end{equation}
where  $F\left(x\right)=\sqrt{g/2}\left[\mathrm{Li}_{2}\left(-e^{-x}\right)-\mathrm{Li}_{2}\left(-e^{x}\right)\right]^{1/2}$, $k_D=4\pi/3a$ and $T_{Fg}=\hbar v\sqrt{\pi n}/k_B$ is the Fermi temperature in monolayer graphene. Over
relatively small ranges of temperatures, as considered in Fig. \ref{fig:DragEqual},
the value of $\beta\mu$ at which $\rho_{D}$ peaks depends weakly
on $T$ and $n$ \cite{Narozhny2012}. Thus, to leading order in $T$ the maximum of the
drag resistivity can be determined by treating $\beta\mu$ as constant,
yielding the relation $v_{\mathrm{max}}^{\star}/v\propto T/\sqrt{n}$.
Eq. \eqref{eq:tmax} is shown in solid and dashed lines Fig. \ref{fig:DragEqual}(c), for fixed values of $\beta\mu$.
The small departure of Eq. \eqref{eq:tmax} from the numerical results is due to the small temperature dependence of the value of $\beta\mu$ at which the drag peaks. From an experimental point of view, the location of the maxima of the drag resistivity can be used to obtain, for example, information about the hopping parameters $w_{0}$ and $w_{1}$.

Fig. \ref{fig:DragEqual}(d) shows that the overall drag behavior, in the case of equal twist angle,
is largely independent of the scattering time within each TBG. Any particular energy dependence of the scattering mechanism appears to mainly modify the magnitude of the drag resistivity, particularly around its peak, but it \textit{does not} modify where such unique peak occurs (around
$\mu/k_BT\sim 2$ \cite{Narozhny2012,Lux2012}). This can be traced to the fact that the drag resistivity comes from
a ratio $\sim\sigma_{D}^{2}/\sigma_{1}\sigma_{2}$ in which all conductivities
depend directly on the scattering time, in such a way that the effect of $\tau_{\mathbf{k}}$ tends to be compensated
in $\rho_{D}$ \cite{Amorim2012,Narozhny2012,Carrega2012}. Note that this is not the case for the drag conductivity, which, as it occurs with the conductivities $\sigma_{1,2}$, it can depend strongly on the scattering mechanism \cite{Yudhistira2019}.

\subsection{Different twist angles}

\begin{figure}[t]
	\includegraphics[scale=0.35]{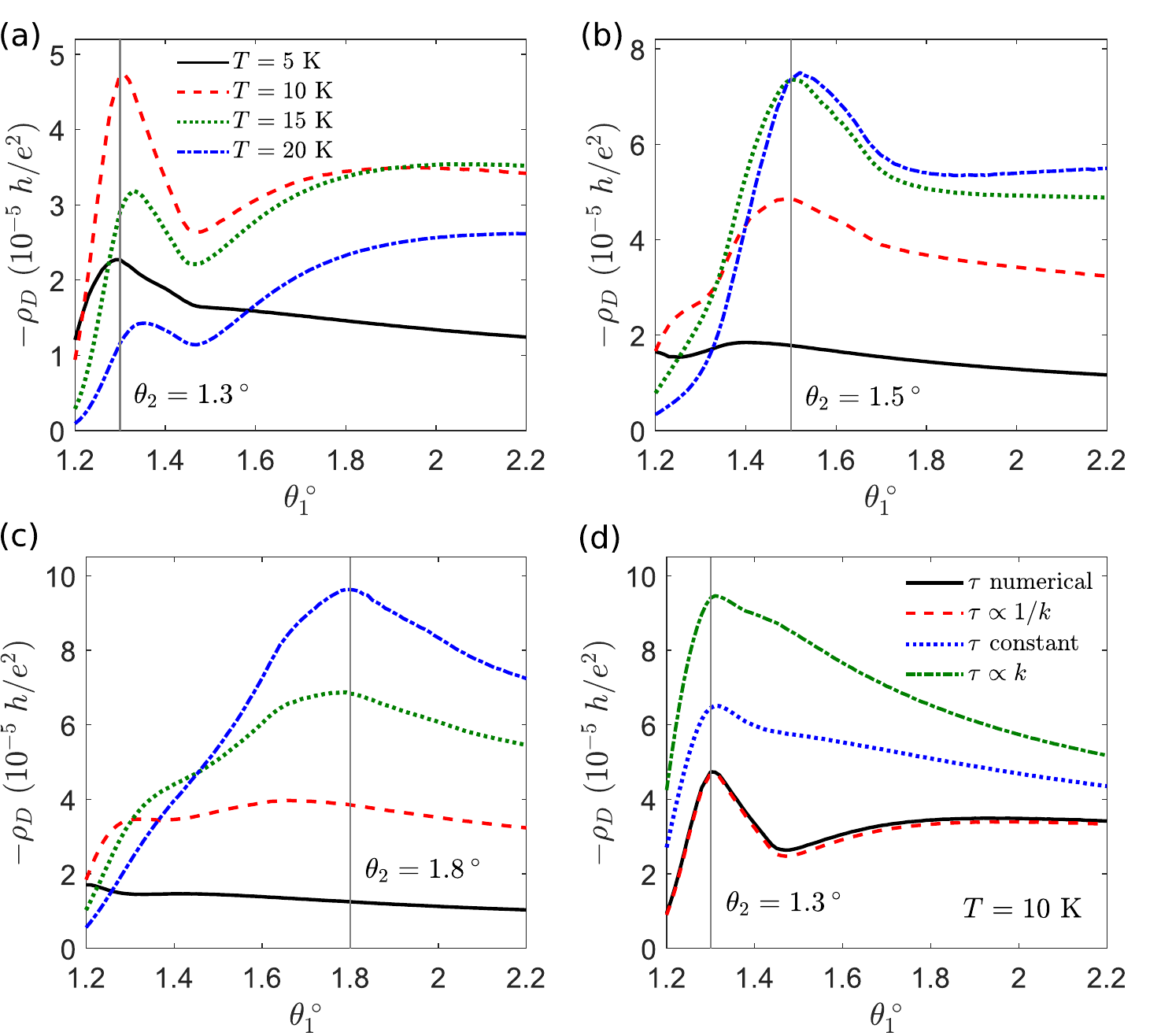}	
	\caption{(a)--(c) Drag resistivity for different twist angles. In each case the twist angle $\theta_2$ in one TBG is fixed (thin vertical line) as the other $\theta_1$ is varied. All other parameters as in Fig. \ref{fig:DragEqual}. (d) Drag resistivity for different scattering times. A non-monotonic behavior appears due to the particular momentum-dependence of $\tau$ in metallic TBG.}\label{fig:DragDiff}
\end{figure}

The behavior of the drag resistivity changes drastically when the
TBG have different twist angle.
 Fig. \ref{fig:DragDiff} shows
$\rho_{D}$ as function of the twist angle $\theta_{1}$ in one TBG,
when the twist angle $\theta_{2}$ in the other TBG is kept fixed.
In general we observe multiple peaks in the drag resistivity, which depend nontrivially on other parameters of the system, such as the temperature, carrier density and interlayer separation. The maximum of $\rho_D$ always occurs around $\theta_1\sim\theta_2$, albeit with a slight shift as the temperature increases. A distinctive minimum in the drag resistivity is seen for $\theta_1>\theta_2$ only when the temperature and the twist angles are relatively low, such that one TBG is at least within a nondegenerate regime. As we discuss in detail below, the observed behavior can be related to an interplay between the dominant scattering mechanisms within each TBG, and the twist-dependence of the response functions $\Gamma_{\ell}$.

\begin{figure}[t]
	\includegraphics[scale=0.34]{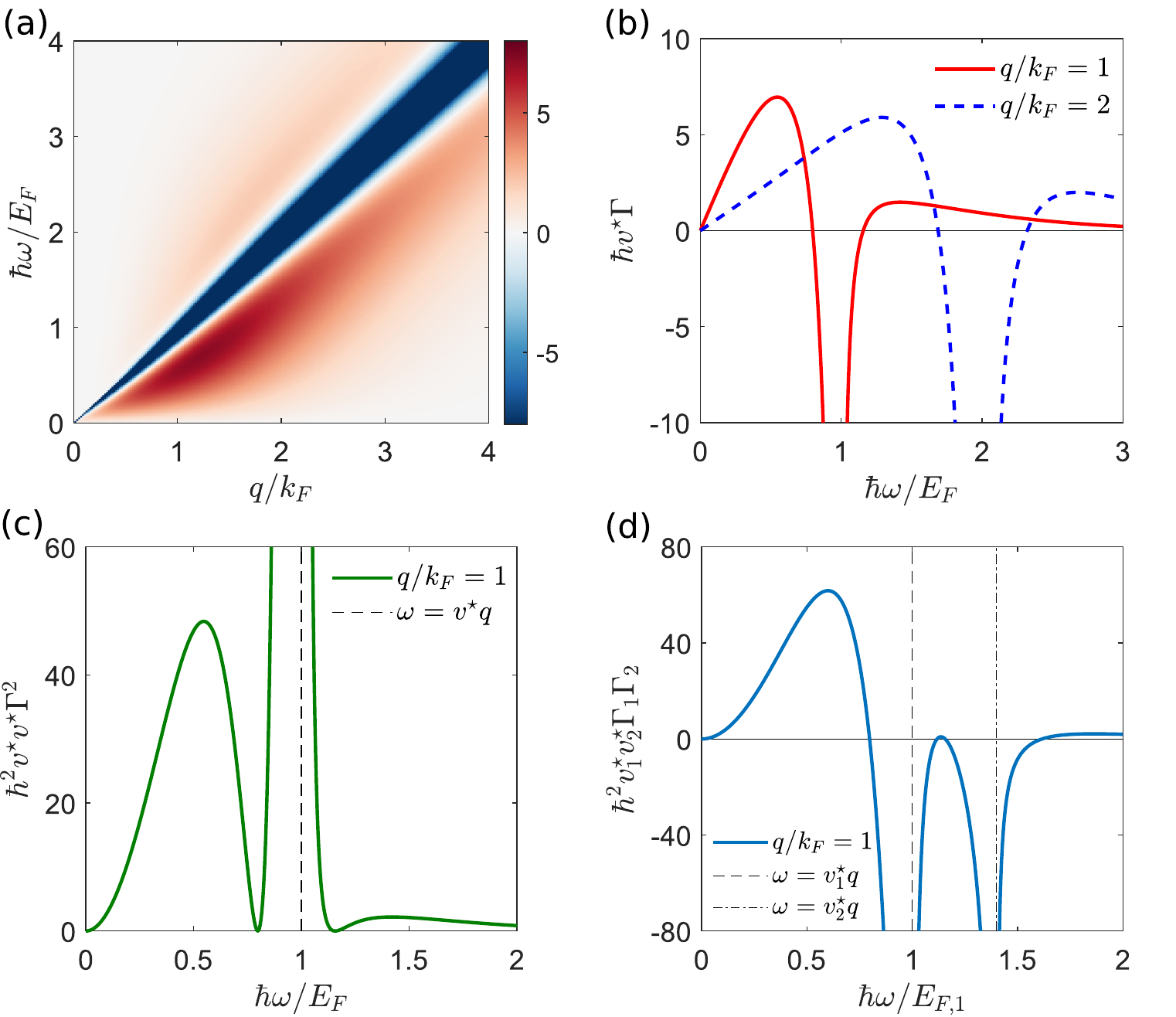}	
	\caption{NLS behavior in TBG, for $\theta=1.3^{\circ}$, $T=10$ K, $n=5\times10^{10}\;\mathrm{cm^{-2}}$ and $n_i=10^{10}\;\mathrm{cm^{-2}}$. Panel (a) shows a density plot of the scaled NLS $\hbar v^{\star}\Gamma$, while (b) shows cuts on the momentum plane. In both cases it can be seen that the NLS changes sign at each side of the line $v^{\star}q=\omega$. (c) and (d) Result of multiplying two NLS with equal and different twist angles, respectively. In the first case the result is always positive-value and peaks only at $v^{\star}q=\omega$. In the second case, the result can be negative at certain regions, and has two distinctive peaks at $v^{\star}_1q=\omega$ and $v^{\star}_2q=\omega$.} \label{fig:NLSTBG}
\end{figure}

The latter is directly reflected in the product 
$\sim\Gamma_{1}\Gamma_{2}$ in Eq. \eqref{eq:dragcond}. Each NLS is a piecewise function that
changes quite abruptly at $v^{\star}_{\ell}q=\omega$, which roughly divides the interband and intraband scattering regimes, cf. Eq. \eqref{eq:NLS}. As a result, the product of two NLS can be markedly different depending on the twist angle in each TBG. In particular, the behavior of $\rho_{D}$ is dictated by which layer has the larger twist angle. Furthermore, it is also strongly influenced by how is the scattering mechanism within each TBG, see Fig. \ref{fig:DragDiff}(d). This signals that the usual compensation of the scattering time in the ratio $\sim\sigma_{D}^{2}/\sigma_{1}\sigma_{2}$ does not take place when the twist angles differ. To understand this behavior, in Fig. \ref{fig:NLSTBG} we show the NLS in TBG, and the product between them as it enters into the drag conductivity kernel. The density plot in Fig. \ref{fig:NLSTBG}(a) shows the characteristic behavior
of the NLS for $\tau\propto1/k$ (see Fig. \ref{fig:Scattering}): it peaks around $v^{\star}q=\omega$, from which it decreases in magnitude as $\left|v^{\star}q-\omega\right|>0$, up until it becomes zero at certain lines along each side of $v^{\star}q=\omega$, from where the NLS then increases and decreases again with opposite sign. This change of sign in the NLS arises due to an inverse energy dependence of the scattering time \cite{Sharma2021} (it does not occur for, e.g., a constant scattering time or $\tau\propto k$). 

The consequences of such behavior in the drag effect can then be intuitively understood by analyzing Figs. \ref{fig:NLSTBG}(c) and \ref{fig:NLSTBG}(d), which show a momentum cut at $q/k_{F}$ of the product of two NLS, in the case of
(c) equal twist angle and (d) different twist angles. In the first
case, since the NLS product $\sim\Gamma^{2}$ is always positive, the
change of sign as each NLS decreases away from $v^{\star}q=\omega$ is only seen as small peaks at each side of it. In contrast, the product
of two NLS with different twist angles yields regions
at which the result is negative, and more importantly, where it changes
sign between its peaks. 
This holds in general, with different weight, for any value
of $q$. Since all other quantities that determine $\sigma_D$
in Eq. \eqref{eq:dragcond} are positive, the net effect of a change of sign in the product $\Gamma_{1}\Gamma_{2}$ is to lower or raise the drag resistivity, depending on the relation
between $\theta_{1}$ and $\theta_{2}$. It is interesting to note that sign reversals in the drag resistivity, without a change of carrier type in each layer, have been measured in electron-hole double bilayer graphene systems \cite{Lee2016,Li2016}, and attributed to a multiband mechanism that can change the sign of the product of two NLS \cite{Zarenia2018}. Here, we emphasize that such behavior is a direct consequence of the momentum-dependence of the scattering
time in TBG.

\section{Conclusions}\label{sec:Conclusions} 

We have studied the Coulomb drag between two metallic TBG, separated so that they only couple through long-range Coulomb interactions. The drag resistivity is calculated taking into account the contributions of gauge phonons and charged impurities to the scattering in TBG. The proposed drag setup assumes that the twist angle in each TBG can be varied independently. In the case of equal twist angles, the drag resistivity follows the expected behavior of exhibiting a unique maximum as the system crossovers from a degenerate to a nondegenerate regime. This crossover can take place solely by varying the twist angle. When the twist angles in each TBG differ, we have found an anomalous drag effect, characterized by the appearance of multiple peaks that depend on the difference between the angles, as well as other parameters of the system. This behavior arises from sign changes in the product of two nonlinear susceptibilities with different twist angles, where due to the momentum dependence of the scattering time in metallic TBG, the result can be negative or positive depending on the difference between the twist angles. Such change of sign influences the magnitude of the drag conductivity and leads to a non-monotonic drag effect.

\begin{acknowledgments}
This paper was partially supported by grants of CONICET (Argentina National Research Council) and Universidad
Nacional del Sur (UNS) and by ANPCyT through PICT 2019-
03491 Res. No. 015/2021, and PIP-CONICET 2021-2023
Grant No. 11220200100941CO. J.S.A. acknowledges support as a member of CONICET, F.E. acknowledges support
from a research fellowship from this institution.
\end{acknowledgments}

\onecolumngrid
\appendix

\section{Nonlinear susceptibility for isotropic scattering time}\label{app:NLS}

In this appendix we give details of the calculation of the NLS given by Eq. \eqref{eq:NLS0}, assuming an isotropic scattering time $\tau_{\ell,\mathbf{k}}=\tau_{\ell}\left(\left|\mathbf{k}\right|\right)$. Following Eq. \eqref{eq:dragcond}, we only
compute the NLS for $\omega>0$. Within the two-band Dirac approximation
of TBG (Sec. \ref{subsec:TwoBand}), we have
\begin{align}
	\epsilon_{\ell,\mathbf{k},s} & =s\hbar v_{\ell}^{\star}\left|\mathbf{k}\right|,\qquad\mathbf{v}_{\ell,\mathbf{k},s}=sv_{\ell}^{\star}\hat{\mathbf{k}},\\
	F_{ss'}\left(\mathbf{k},\mathbf{q}\right) & =\frac{1}{2}\left[1+ss'\frac{k+q\cos\left(\varphi_{\mathbf{k}}-\varphi_{\mathbf{q}}\right)}{\left|\mathbf{k}+\mathbf{q}\right|}\right].
\end{align}
We separate $\boldsymbol{\Gamma}_{\ell}\left(\mathbf{q},\omega\right)=\sum_{s,s'}\boldsymbol{\Gamma}_{\ell,s,s'}\left(\mathbf{q},\omega\right)$.
By introducing the change of angle $\varphi=\varphi_{\mathbf{k}}-\varphi_{\mathbf{q}}$,
the Dirac delta in the NLS can be resolved as
\begin{align}
	\delta\left(\hbar\omega+\epsilon_{\ell,\mathbf{k},s}-\epsilon_{\ell,\mathbf{k}+\mathbf{q},s'}\right) & =\frac{2}{\hbar v_{\ell}^{\star}}\frac{\left|\omega_{\ell}+sk\right|\sum_{i=0,1}\delta\left(\varphi-\varphi_{i}\right)}{\sqrt{\left(q^{2}-\omega_{\ell}^{2}\right)\left[\left(\omega_{\ell}+2sk\right)^{2}-q^{2}\right]}},\label{eq:delta}
\end{align}
where $\varphi_{1}=2\pi-\varphi_{0}$ with $\cos\varphi_{0}=\left(\omega_{\ell}^{2}+2sk\omega_{\ell}-q^{2}\right)/2kq$.
Then in Eq. \eqref{eq:NLS} we have
\begin{equation}
	\left(\tau_{\ell,\mathbf{k}}\mathbf{v}_{\mathbf{k},s}-\tau_{\ell,\mathbf{k}+\mathbf{q}}\mathbf{v}_{\ell,\mathbf{k}+\mathbf{q},s'}\right)\rightarrow-sv_{\ell}^{\star}\frac{\mathbf{q}}{2q^{2}}\frac{1}{k+s\omega_{\ell}}\frac{1}{k}Y_{ss'}\left(q,\omega_{\ell},k\right),
\end{equation}
where
\begin{equation}
	Y_{ss'}\left(q,\omega,k\right) =\tau\left[s'\left(\omega+sk\right)\right]k\left(q^{2}+\omega^{2}+2sk\omega\right)+\tau\left(k\right)\left(k+s\omega\right)\left(q^{2}-\omega^{2}-2sk\omega\right).
\end{equation}
In the above we have used that $\mathbf{k}=\mathbf{q}\left(k/q\right)\cos\varphi+\left(\mathbf{e}_{z}\times\mathbf{q}\right)\left(k/q\right)\sin\varphi$,
and we have dropped terms proportional to $\sin\varphi$ because they
vanish after the integration over the angle [cf. Eq. \eqref{eq:delta}]. Resolving the angle integration by imposing the restrictions $\left|\cos\varphi_{0}\right|<1$
and $\omega>0$, we find
\begin{align}
	\boldsymbol{\Gamma}_{\ell,-,-}\left(\mathbf{q},\omega>0\right) & =-\frac{g}{4\pi\hbar}\Theta\left(q-\omega_{\ell}\right)\frac{\mathbf{q}}{\sqrt{\left|q^{2}-\omega_{\ell}^{2}\right|}}\frac{1}{q^{2}}\int_{\left(q+\omega_{\ell}\right)/2}^{\infty}dkI_{-}\left(q,\omega_{\ell},k\right)Y_{--}\left(q,\omega_{\ell},k\right),\\
	\boldsymbol{\Gamma}_{\ell,-,+}\left(\mathbf{q},\omega>0\right) & =-\frac{g}{4\pi\hbar}\Theta\left(\omega_{\ell}-q\right)\frac{\mathbf{q}}{\sqrt{\left|q^{2}-\omega_{\ell}^{2}\right|}}\frac{1}{q^{2}}\int_{\left(\omega_{\ell}-q\right)/2}^{\left(\omega_{\ell}+q\right)/2}dkI_{-}\left(q,\omega_{\ell},k\right)Y_{-+}\left(q,\omega_{\ell},k\right),\\
	\boldsymbol{\Gamma}_{\ell,+,-}\left(\mathbf{q},\omega>0\right) & =0,\\
	\boldsymbol{\Gamma}_{\ell,+,+}\left(\mathbf{q},\omega>0\right) & =\frac{g}{4\pi\hbar}\Theta\left(q-\omega_{\ell}\right)\frac{\mathbf{q}}{\sqrt{\left|q^{2}-\omega_{\ell}^{2}\right|}}\frac{1}{q^{2}}\int_{\left(q-\omega_{\ell}\right)/2}^{\infty}dkI_{+}\left(q,\omega_{\ell},k\right)Y_{++}\left(q,\omega_{\ell},k\right),
\end{align}
where
\begin{equation}
	I_{\pm}\left(q,\omega,k\right)=\left[f_{\ell}\left(\pm k\right)-f_{\ell}\left(\omega\pm k\right)\right]\sqrt{\left|\left(\omega\pm2k\right)^{2}-q^{2}\right|}\frac{1}{k\left(k\pm\omega\right)}.
\end{equation}
From here Eq. \eqref{eq:NLS} follows after rewriting the $k$-integrals by changing variables and regrouping terms. 

\section{Dynamical screening and collinear singularity}\label{app:Screening}

The collinear singularity, within the Dirac approximation, gives rise
to the divergences in the NLS when $v_{\ell}^{\star}q\rightarrow\omega$
\cite{Narozhny2016}. Here we show that these divergences are cured in
the calculation of the drag conductivity when the dynamical screening
of the interlayer interaction is taken into account \cite{Carrega2012}. We start
by writing the effective dielectric function \eqref{eq:dielectricF} as
\begin{align}
	\varepsilon_{12}\left(q,\omega\right) & =\left(1+\frac{q_{1}}{q}\right)\left(1+\frac{q_{2}}{q}\right)-\frac{q_{1}q_{2}}{q^{2}}e^{-2qd}.
\end{align}
Here we defined an effective Thomas-Fermi (TF) wave vector $q_{\ell}=q_{R,\ell}+iq_{I,\ell}$
in the TBG $\ell=1,2$, where $q_{R/I,\ell}=q_{0T,\ell}\tilde{\Pi}_{R/I,\ell}$
with $q_{0T,i}=g\alpha_{\ell}k_{F,\ell}$ being the zero temperature
TF vector, and $\tilde{\Pi}_{R/I,\ell}=\Pi_{R/I,\ell}/D_{0}$ the
real and imaginary parts of the polarization function \eqref{eq:pol}, scaled
by the density of states $D_{0}=gk_{F,\ell}/2\pi\hbar v^{\star}_{\ell}$.
To compute $\tilde{\Pi}_{R}$ and $\tilde{\Pi}_{I}$ we use the semi-analytical
expressions of Ref. \cite{Ramezanali2009}, which for $\omega>0$ can be written as
\begin{align}
	\tilde{\Pi}_{R,\ell} & =\frac{k_{B}T}{E_{F,\ell}}\mathcal{F}_{\ell}-\frac{q}{4k_{F,\ell}}\frac{1}{\sqrt{\left|1-\omega_{\ell}^{2}/q^{2}\right|}}\left[\Theta\left(\omega_{\ell}-q\right)\mathcal{G}_{\ell}\left(q,\omega,T\right)+\Theta\left(q-\omega_{\ell}\right)\mathcal{H}_{\ell}\left(q,\omega,T\right)\right],\\
	\tilde{\Pi}_{I,\ell} & =\frac{q}{4k_{F,\ell}}\frac{1}{\sqrt{\left|1-\omega_{\ell}^{2}/q^{2}\right|}}\left[\Theta\left(q-\omega_{\ell}\right)\mathcal{G}_{\ell}\left(q,\omega,T\right)-\Theta\left(\omega_{\ell}-q\right)\mathcal{H}_{\ell}\left(q,\omega,T\right)\right],
\end{align}
where $\mathcal{F}_{\ell}=\ln\left[2\left(1+\cosh\beta\mu_{\ell}\right)\right]$ and
\begin{align}
	\mathcal{G}_{\ell}\left(q,\omega,T\right) & =\sum_{a,b=\pm1}\int_{1}^{\infty}du\frac{a\sqrt{u^{2}-1}}{\exp\left[\left(\hbar\left|v_{\ell}^{\star}qu-a\omega\right|+2b\mu_{\ell}\right)/2k_{B}T\right]+1},\\
	\mathcal{H}_{\ell}\left(q,\omega,T\right) & =-\frac{\pi}{2}+\sum_{a,b=\pm1}\int_{0}^{1}du\frac{\sqrt{1-u^{2}}}{\exp\left[\left(\hbar\left|v_{\ell}^{\star}qu-a\omega\right|+2b\mu_{\ell}\right)/2k_{B}T\right]+1}.
\end{align}
Now we redefine
\begin{align}
	\tilde{\Pi}_{R/I,\ell} & \rightarrow\frac{1}{\sqrt{\left|1-\omega_{\ell}^{2}/q^{2}\right|}}\tilde{\Pi}_{R/I,\ell},
\end{align}
which in turn implies $q_{R/I,\ell}\rightarrow\tilde{q}_{R/I,\ell}\left[1-\omega_{\ell}^{2}/q^{2}\right]^{-1/2}$.
The dielectric function can then be written as 
\begin{equation}
	\varepsilon_{12}\left(q,\omega\right)=\frac{\tilde{\varepsilon}_{12}\left(q,\omega\right)}{q^{2}\sqrt{\left|1-\omega_{1}^{2}/q^{2}\right|\left|1-\omega_{2}^{2}/q^{2}\right|}},\label{eq:e12}
\end{equation}
where
\begin{equation}
	\tilde{\varepsilon}_{12}\left(q,\omega\right)=\left(q\sqrt{\left|1-\omega_{1}^{2}/q^{2}\right|}+\tilde{q}_{1}\right)\left(q\sqrt{\left|1-\omega_{2}^{2}/q^{2}\right|}+\tilde{q}_{2}\right)-\tilde{q}_{1}\tilde{q}_{2}e^{-2qd},
\end{equation}
with $\tilde{q}_{\ell}=\tilde{q}_{R,\ell}+i\tilde{q}_{I,\ell}$.
The divergences now only appear in the denominator of Eq. \eqref{eq:e12}. By
replacing $\varepsilon_{12}\left(q,\omega\right)$ above, and the projected
NLS given by Eq. \eqref{eq:NLS} (choosing the current in the $x$ axis), the
drag conductivity \eqref{eq:dragcond} becomes
\begin{equation}
	\sigma_{D}=\frac{e^{2}}{h}\frac{g^{2}}{64\pi}\frac{\hbar}{k_{B}T}\alpha_{1}^{\star}\alpha_{2}^{\star}v_{1}^{\star}v_{2}^{\star}\int dq\frac{e^{-2qd}}{q}\int_{0}^{\infty}d\omega\frac{\sqrt{\left|1-\omega_{1}^{2}/q^{2}\right|}\sqrt{\left|1-\omega_{2}^{2}/q^{2}\right|}}{\left|\tilde{\varepsilon}_{12}\left(q,\omega\right)\right|^{2}}\frac{\Gamma_{1}\left(q,\omega\right)\Gamma_{2}\left(q,\omega\right)}{\sinh^{2}\left(\hbar\omega/2k_BT\right)},\label{eq:dragcond2}
\end{equation}
where
\begin{equation}
	\Gamma_{\ell}\left(q,\omega\right)=\Theta\left(q-\omega_{\ell}\right)\Gamma_{\ell,+}\left(q,\omega\right)+\Theta\left(\omega_{\ell}-q\right)\Gamma_{\ell,-}\left(q,\omega\right).
\end{equation}
The expression \eqref{eq:dragcond2} explicitly shows that the divergences in both the dielectric function and the NLS, when $v^{\star}_{\ell}q\rightarrow\omega$,
are effectively cured in the integral kernel.

\twocolumngrid

\end{document}